\documentclass[secnumarabic,amssymb, nobibnotes, aps, prb,preprint]{revtex4-1}
\usepackage{graphicx}
\usepackage{hyperref}
\usepackage{dcolumn}
\draft 

\begin{document}


\title{If graphynes turn into graphene: the thermal stability study} 



\author{Yi-Guo Xu}
\affiliation{Institute of Modern Physics, Fudan University, Shanghai 200433, China}
\affiliation{Applied Ion Beam Physics Laboratory, Key Laboratory of the Ministry of Education, Fudan University, Shanghai 200433, China}

\author{Chen Ming}
\affiliation{Institute of Modern Physics, Fudan University, Shanghai 200433, China}
\affiliation{Applied Ion Beam Physics Laboratory, Key Laboratory of the Ministry of Education, Fudan University, Shanghai 200433, China}

\author{Zheng-Zhe Lin}
\affiliation{Institute of Modern Physics, Fudan University, Shanghai 200433, China}
\affiliation{Applied Ion Beam Physics Laboratory, Key Laboratory of the Ministry of Education, Fudan University, Shanghai 200433, China}

\author{Fan-Xin Meng}
\affiliation{Institute of Modern Physics, Fudan University, Shanghai 200433, China}
\affiliation{Applied Ion Beam Physics Laboratory, Key Laboratory of the Ministry of Education, Fudan University, Shanghai 200433, China}

\author{Jun Zhuang}
\affiliation{Department of Optical Science and Engineering, Fudan University, Shanghai 200433, China}

\author{Xi-Jing Ning}
\email[]{xjning@fudan.edu.cn}
\affiliation{Institute of Modern Physics, Fudan University, Shanghai 200433, China}
\affiliation{Applied Ion Beam Physics Laboratory, Key Laboratory of the Ministry of Education, Fudan University, Shanghai 200433, China}


\date{\today}

\begin{abstract}
The thermal stability of $\alpha$-, $\beta$-, 6,6,12-graphyne and graphdiyne was studied by a statistic model, which was seriously tested by classical molecular dynamics simulations. By first-principles calculations of related potential energy curves, the model predicts that all the lifetime of free-standing single layer graphynes considered is more than 10$^{44}$ years at room temperature. When the temperature gets up to 1000 K, they are still very stable, but quickly turn into graphene if the temperature is about 2000 K.
\end{abstract}


\maketitle 

\section{Introduction}
\begin{figure}
 \includegraphics[width=8.5cm]{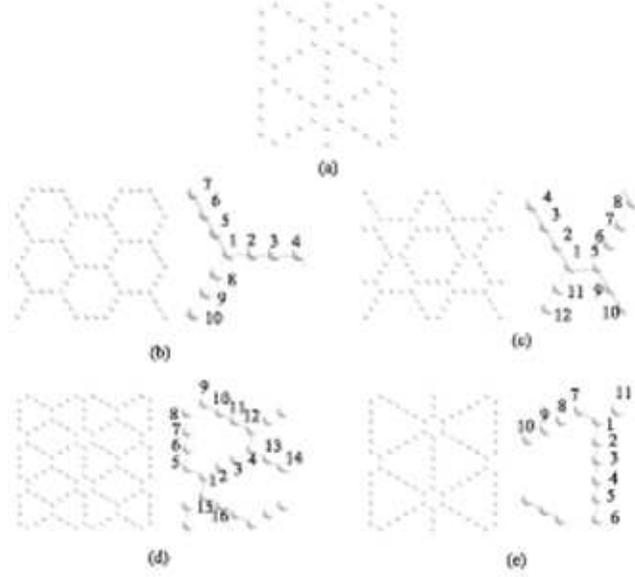} %
 \caption{Optimized geometrical structure of (a) graphyne, (b) $\alpha$-graphyne, (c) $\beta$-graphyne, (d) 6,6,12-graphynes and (e) graphdiyne.\label{Fig1}}%
 \end{figure}

As a fundamental element of life on earth, carbon can form numerous carbon allotropes consisting of three hybridization state (\emph{sp, sp$^{2}$, sp$^{3}$}), such as graphite, diamonds, fullerene, carbon nanotubes and graphene.\cite{1985kroto162Nat,1991iijima56Nat,2004norskov666Sci} Due to the distinct mechanical and physical properties, carbon allotropes have attracted tremendous attention, especially since the successful preparation of graphene. Because of its single-layer structure, graphene is proposed as a promising candidate material for tinier and higher frequency transistor to replace silicon. However, graphene does not possess a band gap.\cite{2005norskov197Nat} Even if chemical or electrical doping is applied, the gap can be only opened by a few meV,\cite{2011liu3335} which fails to meet the requirement of realistic applications. Graphyne, another allotrope of carbon, was first predicted by Baughman et al. in 1987.\cite{1987baughman6687JCP} As shown in Fig. \ref{Fig1}, this allotrope can be grouped as $\alpha$-, $\beta$-, $\gamma$- and 6,6,12-graphyne. Replacing the acetylenic linkages (--C$\equiv$C--) in $\gamma$-graphyne with diacetylenic linkages (--C$\equiv$C--C$\equiv$C--) yields a new structure of carbon allotropes, graphdiyne, which was proposed by Haley et al. in 1997.\cite{1997haley836ACIE} Among these structures, graphdiyne and $\gamma$-graphyne are predicted to own a bandgap of 0.52 eV and 0.53 eV ,respectively.\cite{1998narita11009PRB}

In the last decade, tremendous efforts have been made for the preparation of graphynes, but only some precursors and subunits of graphynes were synthesized.\cite{1997haley836ACIE,1997haley2956JACS,2000kehoe969,2001wan3893,2003marsden2355,2005marsden10213,
2006yoshimuras2933,2007johnson3725,2007gholami9081ACIE,2008haley519,2010diederich803AM,2010liu1496} In 2010, Li et al. reported that thin films of graphdiyne were successfully synthesized on copper surfaces and the film shows semiconductive properties.\cite{2010li3256} Shortly afterwards, the same group synthesized graphdiyne nanotube arrays and graphdiyne nanowires.\cite{2012qian730,2011li2611JPCC} Triggered by these experiments, a series of theoretical works have ben made to predict the properties of graphynes in recent years. For example, first-principles calculations indicate that the bandgap of graphdiyne is about 1.2 eV,\cite{2011jiao11843CC,2012bu3934JPCA} which is comparable to silicon, while the bandgap of graphyne is about 0.96 $eV$ and both can be modulated.\cite{2012bu3934JPCA,2011kang20466JPCC} The other graphynes, such as $\alpha$-, $\beta$-,6,6,12-graphyne are predicted to have Dirac cones by first-principles calculation, and the cone of 6,6,12-graphyne is directionally anisotropic and nonequivalent,\cite{2012kim115435PRB} which is more versatile than graphene.\cite{2012malko086804PRL} In addition, a lots of calculations show that graphynes can be used for hydrogen storage and gas separation.\cite{2011jiao11843CC,2011zhang8845JPCC,2012guo13837JPCC,2012srinivasu5951JPCC}

It is notable that synthesis of large homogenous sheets of single-layer graphyne have not yet been reported up to now and first-principles calculations show graphyne is less stable than graphene because the binding energy of graphyne is lower than graphene by about 0.56 eV/atom.\cite{2011zhang8845JPCC} Accordingly, it remains uncertain if free-standing graphyne can survive at room temperature, and how long it will survive before they turn into graphene at higher temperatures. These questions challenge current theories concerning the lifetime of nanodevices, the fatigue time and creep rate of bulk material and so on.\cite{1997zhang377Sci,2007mei1175} Molecular dynamics (MD) simulations seem direct solutions to this problem, but lifetime of most materials at room temperature is far beyond the time scale of MD simulations, which can cover only several microseconds at most.

Very recently, a simple model based on the statistic of individual atoms was developed and has been successfully applied to predict the lifetime of carbon monatomic chains and single wall carbon nanocones,\cite{2011lin40002EPL,2012ming2651Carbon} and has been extended to predict thermal reaction rate.\cite{2012li08504CPL} In this work, this model was applied to predict the thermal stability of $\alpha$-, $\beta$-, 6,6,12-graphyne and graphdiyne. Firstly, MD simulations with the empirical potential for C-C interaction were performed at higher temperatures to explore all possible paths for $\alpha$-graphyne to turn into graphene, showing that several defects formed in graphyne will lead to formation of a hexagon, a subunit of graphene, and then avalanches of dislocation take place around the ring, and finally the structure of graphene forms. So the time taken by the formation of the primary defects should be the lifetime of graphynes. Secondly, we obtained the the time taken by defect formation at higher temperatures separately by a great deal of MD simulations and by our statistic model with the same empirical potential in order to test the accuracy of our model. To apply our model, we calculated the potential energy curve (PEC) along minimum energy path (MEP) for the defect formation and used these data in the model to predict the time taken by defect formation. The results are in good agreement with the MD simulations. As a comparison, the harmonic transition state theory was also applied, but the results are far from the MD simulations. Finally, for accurate prediction, first-principles calculations were performed to obtain the PEC along MEP for the defect formation for all the graphynes, and then predicted their lifetime at any temperatures by our statistic model.

\section{Methods}
\begin{figure}
 \includegraphics[width=8.5cm]{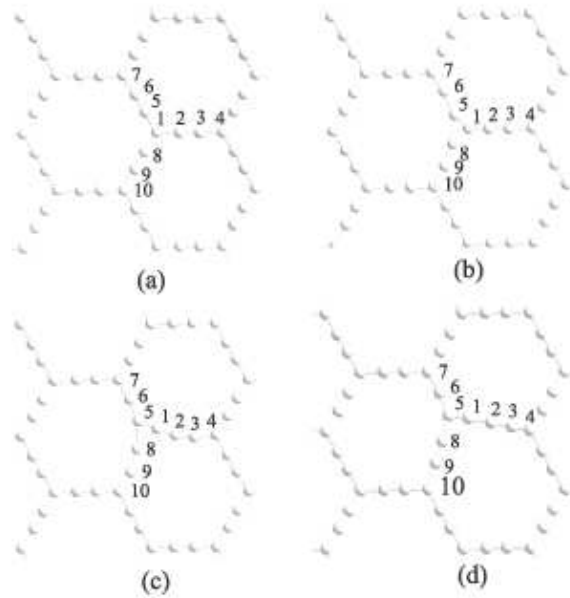} %
 \caption{The defect formation process of $\alpha$-graphyne drawn from the MD simulations.\label{Fig2}}%
 \end{figure}

The statistic model is based on the fact that the kinetic energy ($\varepsilon$) of a single atom in condensed matters obeys the Boltzmann distribution,$e^{-1/2} e^{-\varepsilon /k_B T}$ , which has already been confirmed by a great deal of MD simulations.\cite{2011lin40002EPL,2012ming2651Carbon,2010han064103JCP}Therefore, within a time unit, the total time for an atom to obtain a kinetic energy larger than the barrier  is\cite{2011lin40002EPL}	
\begin{equation}  \label{equ1}
  t=Z^{-1}\int_{E_0}^\infty  e^{-1/2} e^{-\varepsilon /k_B T} d\varepsilon ,
\end{equation}
where $Z=\frac{\sqrt{\pi}}{2} {(k_B T)}^{\frac{3}{2}}$  is the partition function. Considering an atom located at the bottom of a potential well $V(x)$  with kinetic energy $\varepsilon(\varepsilon > E_0)$, the time taken by this atom to escape from the valley is $\delta t=\sqrt{m}\int_0^a dx/\sqrt{2(\varepsilon-V(x))}$, where  $a$ is the half width of the well, and the average time at a certain temperature $T$  can be obtained by	 	
\begin{equation}\label{equ2}
\overline{\delta t}=\frac{\int_{E_0}^\infty (\delta t) e^{-1/2} e^{-\varepsilon /k_B T}  d\varepsilon}
                          {\int_{E_0}^\infty e^{-1/2} e^{-\varepsilon /k_B T} d\varepsilon}  ,
\end{equation}
So the frequency (or rate) of the hopping event is	
\begin{equation} \label{equ3}
  F=\frac{t}{\overline{\delta t}}=
  \frac{1}{Z}
    \frac{(\int_{E_0}^\infty e^{-1/2} e^{-\varepsilon /k_B T} d\varepsilon)^2}
       {\int_{E_0}^\infty (\delta t) e^{-1/2} e^{-\varepsilon /k_B T}  d\varepsilon}
\end{equation}
Clearly, as long as the PEC for the hopping atom along the MEP is known, the time for this atom staying within the potential well can be obtained as $1/F$. As an example, Fig. \ref{Fig2} displays a possible path for $\alpha$-graphyne turning into graphene, drawn from MD simulations (see below for details). Primarily, the atom chain C1-2-3-4, C1-5-6-7 and C1-8-9-10 form the three arms of the perfect $\alpha$-graphyne, and then C5 and C8 get bonded due to thermal motion [Fig. \ref{Fig2}(a)]. Finally, the bond connecting atom C1 and atom C8 breaks, producing a long atomic chain consisting of C5-1-2-3-4 and a short one composed of C5-6-7. This process can be regarded as a hopping event by atom C8 crossing over a barrier $E_0$ . Once the PEC along the MEP is determined, the time for C5-C8 bonding can be obtained by $1/F$.
The MD model for simulating the evolution of graphyne at temperature   consists of a piece of single layer graphyne with periodic boundary conditions imposed on the plane. The interaction between carbon atoms were described by Brenner potential,\cite{990brenner9458PRB} and the velocity Verlet algorithm was used with a time step of 0.2 fs. The constant temperature is realized by selecting one atom randomly every 20 fs to reset the velocity $( v_i ^{old})$ of an atom   as\cite{1988riley5934JCP}	
\begin{equation}\label{equ4}
  v_i ^{new}=(1-\theta)^{1/2} v_i ^{old}+ \theta^{1/2} v_i ^T (\xi) ,(i=x,y,z)
\end{equation}
where $ v_i ^T (\xi) $is a random velocity chosen from the Maxwellian distribution at temperature $T$, and   $\theta (0<\theta\leq 1)$ is a random number controlling the reset.

For accurate calculation, we performed first-principles calculations on the structure of single layer graphynes and the PEC for the defect formation via generalized gradient approximation (GGA) with Perdew-Burke-Ernzerhof (PBE) exchange correlation functional implemented in DMol3 package.\cite{1990delley508JCP,2000delley122CMS,1996perdew3865PRL} A 2$\times$2$\times$1 supercell was used to simulate infinite single-layer sheet with two-dimensional periodic boundary conditions applied, while a vacuum of 25 \AA\ was applied in the direction perpendicular to the graphyne plane to exclude the interactions between adjacent layers. The K points sampling in Brillouin zone was 4$\times$4$\times$1 generated by the Monkhorst-Pack scheme.\cite{1976hendrik5188PRB} The convergence tolerance of the energy was set to $10^-5$ Ha, and the maximum allowed force and displacement were 0.002 Ha/\AA\  and 0.005 \AA\, respectively. To get the PEC along the MEP, the linear or quadratic synchronous transit (LST/QST) method combined with conjugate gradient refinements was adopted for the transition state (TS) search.\cite{1977halgren225CPL} To And then the nudged elastic band (NEB) method was performed to confirm the transition state connects to the relevant reactant and product.\cite{2000henkelman9978JCP}

\section{Results and Discussions}
\begin{figure}
 \includegraphics[width=8.5cm]{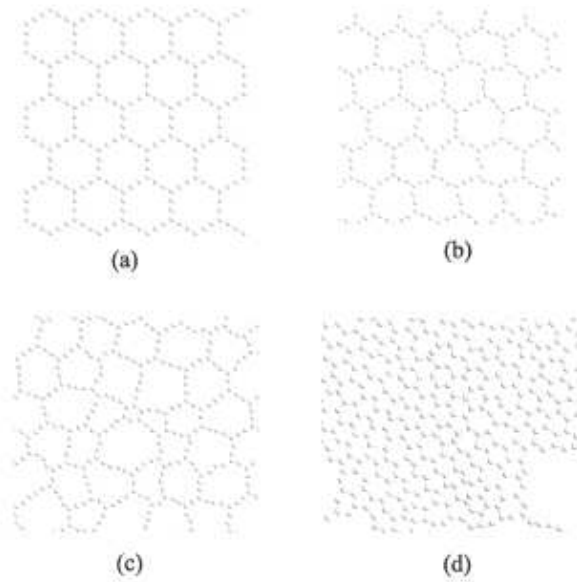} %
 \caption{Schematic of the full atomistic specimen drawn from MD simulations consisting 6.1 nm by 5.9 nm graphyne sheet (720 carbon atoms). (a) perfect $\alpha$-graphyne, (b) defect formed, (c) sixfold carbon ring formed and (d) graphene with defect formed.\label{Fig3}}%
 \end{figure}

To find the turning processes form $\alpha$-graphyne to graphene, MD simulations were performed on the evolution of an $\alpha$-graphyne sheet consisting 720 carbon atoms [Fig. \ref{Fig3}(a)]. Figs. \ref{Fig3}(b)-(d) show a possible way that a perfect $\alpha$-graphyne sheet turns into graphene sheet at 1600 K. After the graphyne relaxing at for about 1 picosecond, an obvious defect begins to form [Fig. \ref{Fig2}], where C5 and C8 get bonded due to thermal motion, and then the bond connecting atom C1 and atom C8 break, producing a long atomic chain consisting of C5-1-2-3-4 and a short one composed of C5-6-7 [Fig. \ref{Fig3}(b)]. As the longer atomic chain continuously gets even longer, a hexagon forms at about 50 ps, as the arrow pointed in Fig. \ref{Fig3}(c). Then avalanches of dislocations take place around this hexagon at about 200 ps, producing the structure of graphene [Fig. \ref{Fig3}(d)]. We noted that graphene formation in MD simulations of other graphynes takes the similar way, i.e, longer atom chains of more than four atoms form firstly, then hexagon form, and finally avalanches of dislocation take place. The time taken by the third process is about 100 times the second process, which is about 10 times longer than the primary defect formation. So the time taken by the formation of the primary defects can be considered as the shortest lifetime of $\alpha$-graphyne.
To test the model strictly, we compared the time taken by the defect formation in $\alpha$-graphyne separately by a great deal of MD simulations and by our statistic model in the temperature range of 1000 to 1600 K. To use the statistic model, the PEC for the defect formation (Fig. \ref{Fig2}) along the MEP was obtained by NEB method,\cite{2000henkelman9978JCP} starting from the initial configuration [Fig. \ref{Fig2}(a)] to the final configuration [Fig. \ref{Fig2}(d)]. The isomerization process obtained by NEB method is very similar to the one of MD simulation. Although several atoms are involved in the process, the binding of C5 and C8 is the key step for this reaction [Fig. \ref{Fig2}(b)]. As the coordinate of C8 atom changes in the isomerization process, the total potential against the coordinate of C8 produces a PEC, which was used in Eq. \ref{equ3} for calculating $\delta t$ and further $F$  was determined by Eq. \ref{equ3}.

For convenient discussion, we denote the atom, such as C1, at the joint of the three atomic chains in perfect $\alpha$-graphyne [Fig. \ref{Fig1}(b)] as the node atom. It is notable that the defect may form at every node of $\alpha$-graphyne, and three atoms, such as C2, C5 and C8, adjacent to a given node atom (C1) are equivalent for the defect formation because anyone can bond with one of the other two atoms to form the defect. So for an $\alpha$-graphyne sheet of $N$  nodes, the total rate of defect formation seems to be $6NF$ . However, we must see the fact that when the C2 atom has a tendency to bond with C5 or C8, the probability for C5 (or C8) independently forming the defect will vanish. Thus, the total rate should be $2NF$  instead of $6NF$  . Another important fact is that the result of total rate  stems from the fact that the movement of every node in $\alpha$-graphyne for defect formation is independent. But this is not the case in realistic system. The movement of every node is not completely independent due to the correlation between the neighboring nodes. So the effective node  will be smaller than the node number $N$, then the real total rate of defect formation should be $2N'F$  and the average rate for single node $P_N$ should be $\frac{2N'F}{N}$ , which is smaller than $2F$ , here the ratio  $\frac{N'}{N}$ can reflect strength of the correlation. Obviously, the correlation between two nodes gets weaker with their distance increasing, and disappears at a cut-off distance. Therefore, the correlation is stronger in a small MD simulation box than that in a large MD simulation box because of the periodic boundary condition, so the ratio  $\frac{N'}{N}$  is smaller in a small box. According to above discussion, we can draw a conclusion that the average rate for single node $P_N$ will increase with the number of node $N$  until it reaches the convergence. Accordingly, for seriously comparing the statistic model with the MD simulation result, we need to obtain the total rate drawn from MD simulations for infinite large graphyne, which was implemented as follows: giving a temperature 1200 $ K$, we repeated MD simulations for given node number of   until the time taken by formation of the defect varies below 1 \%; increasing the node number $N$ , step by step, performing similar MD simulations, and finally, extracting the rate for infinite lager graphyne. As shown in Fig. 4, the average rate for single node indeed increases with the number of the node, and the dependence on the number can be well fitted by $P_N=A-\frac{B}{N}$ , here A and B are two constants. Thus, if a rate $R_N(T)$  is drawn from the MD simulations involving $N$ nodes at temperature $T$, then the rate $R_\infty(T)$ for an infinite sheet at the same temperature should be $R_\infty(T)=\frac{A}{P_N}\cdot R_N(T)$ . Based on this fact, MD simulations involving 60 nodes in temperature range of 1000 to 1600 K were repeated at the temperature points until the formation time of the defect varies below 1 \%, and multiplied the rate for single node by $\frac{A}{P_N}$ . As shown in Fig. \ref{Fig5}, the result of MD simulations is in good agreement with the statistic model prediction.
 \begin{figure}
 \includegraphics[width=8.5cm]{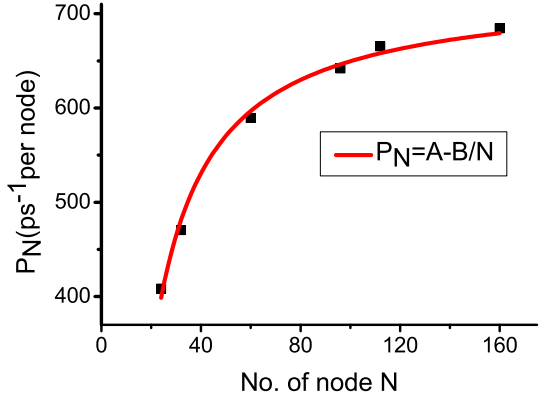} %
 \caption{The rate of defect formation in $\alpha$-graphyne at 1200 K for single node $P_N$ , as a function of nodes number $N$  fitted by a function $P_N=A-\frac{B}{A}$. \label{Fig4}}%
 \end{figure}
  \begin{figure}
 \includegraphics[width=8.5cm]{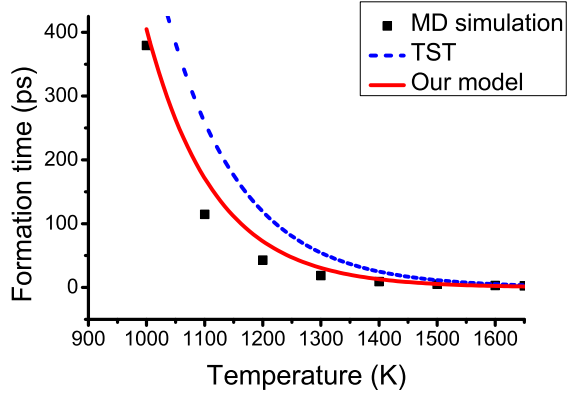} %
 \caption{The average time $t$  of the defect formation in $\alpha$-graphyne of 240 atoms obtained by our statistic model, by TST and by MD simulations.\label{Fig5}}%
 \end{figure}
In principle, conventional transition state theory (TST) is applicable to the above issue. As comparison, harmonic TST was applied to calculate the rate by
\begin{equation}\label{equ5}
  F=\frac{\prod\limits_{i=1}^{3N-6} v_0(i)}{\prod\limits_{i=1}^{3N-5} v_s(i)} e^{E_0/k_B T}
\end{equation}	
where $v_s(i) $ and  $ v_0(i)$ are the frequencies of normal mode around the saddle point and the potential well. As shown in Fig. \ref{Fig6}, the results are a little away from the MD simulations.

Then the model was used for predicting the stability of all the graphynes. Considering the empirical potential may be too rough, first-principles calculations were performed to calculate the potential along the MEP for the defect formation for all graphynes. It is well known that due to the existence of acetylenic linkages (--C$\equiv$C--) in graphynes, bonds of the \textit{sp}-hybridized C atoms along the C-chain are unsaturated and the atoms are chemically active. In $\alpha$-graphyne [Fig. \ref{Fig2}(a)] all the atoms except for the node ones (such as C1) are chemically active. So two of the three atoms C2,C5 and C8 may bond together due to thermal fluctuation, and then one of the bond C1-C5 (or C1-C8 ) break, producing the defect shown in Fig. \ref{Fig2}(d). Thus, in calculation of the PEC for $\alpha$-graphyne by NEB method, the configuration Fig. \ref{Fig2}(a) and Fig. \ref{Fig2}(d) should be taken as the initial and final ones, respectively. However, the first-principles optimization shows that the configuration of Fig.\ref{Fig2}(d) is a transition state and turns into a stable one shown in Fig. \ref{Fig6}(a), which was taken as the final configuration to get the PEC, i.e., the total potential against the coordinate of C8 atom. Using this curve in Eq. \ref{equ3}, the average time for one defect formation is predicted to be $ 1.6\times 10^{79}$ years at room temperature, i.e, the lifetime of $\alpha$-graphyne is of the order of about $1.6 \times 10^{79}$ years at room temperature. For higher temperature, up to 1000 K, the lifetime of $\alpha$-graphyne is also long. When the temperature gets up to 2000 K, the lifetime decreases to 2 hours (Tab. \ref{tab1}).
  \begin{figure}
 \includegraphics[width=8.5cm]{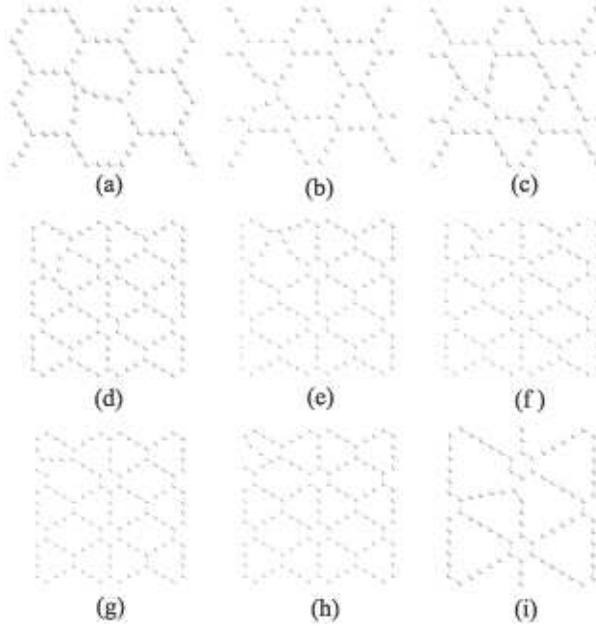} %
 \caption{Optimized geometrical structure of possible defect in (a) $\alpha$-graphyne, (b-c) $\beta$-graphyne, (d-h) 6,6,12-graphynes and (i)  graphdiyne.\label{Fig6}}%
 \end{figure}
 \begin{table}
 \caption{The lifetime of $\alpha$-, $\beta$-, 6,6,12-graphyne and graphdiyne at 300 K, 500 K, 1000 K and the corresponding energy barrier.\label{tab1} }
 \begin{tabular}{cccccc}
   \hline
     & $\alpha$-graphyne & $\beta$-graphyne & 6,6,12-graphynes & graphdiyne  \\
    \hline
   $E_b$(eV)     & 5.83                 & 4,35                 & 4.32                 & 3.72  \\
   300 K (year)  & 1.6 $\times 10^{79}$ & 8.2 $\times 10^{54}$ & 2.4 $\times 10^{54}$ & 5.6 $\times 10^{44}$  \\
   500 K (year)  & 1.5 $\times 10^{40}$ & 6.2 $\times 10^{25}$ & 3.0 $\times 10^{25}$ & 7.1 $\times 10^{19}$  \\
   1000 K (year) & 1.6 $\times 10^{10}$ & 1.1 $\times 10^{4}$  & 7.4 $\times 10^{3}$ & 17.7  \\
   2000 K (s)    & 8518                 & 5                    & 4                    & 0.3  \\
   \hline
 \end{tabular}
\end{table}
In $\beta$-graphyne [Fig.\ref{Fig1}(c)], the atom along acetylenic linkages (--C$\equiv$C--) such as C2 and C3 (or C6,C7,C9,C10,C11,C12) are unsaturated. So C9 may bond together with C11 or C6 by Path I: C9 bond with C11 and then C5-C9 or C1-C11 bond break to form defect shown by Fig. \ref{Fig6}(b); or by Path II: C9 bond with C6 and then C5-C6 or C5-C9 bond break to form defect shown by Fig. \ref{Fig6}(c). The PEC for the two paths were obtained by NEB method, and the potential barrier for path II is 6.03 eV, which is significantly larger than the one for path I, 4.35 eV. So the main defect path of $\beta$-graphyne should be path I, and the corresponding lifetime is about $8\times 10^{54}$ years for room temperature. Even if $\beta$-graphyne is heated up to 1000 K, the lifetime is still as long as 10000 years (Tab. \ref{tab1}).

Compared with the other graphynes, 6,6,12-graphyne [Fig.\ref{Fig1}(d)] has worse symmetry. The acetylenic linkages (C5--C6$\equiv$C7--C8) connecting two hexagons is not exactly equivalent to the C chain (C1--C2$\equiv$C3--C4). So defects may form by five possible paths: Path I: C2 bond with C16 and then C1-C2 or C15-C16 bond break to form defect shown by Fig. \ref{Fig6}(d); Path II: C3 bond with C11 and then C3-C4 or C11-C12 bond break to form defect shown by Fig. \ref{Fig6}(e); Path III: C3 bond with C13 and then C4-C3 or C4-C13 bond break to form defect shown by Fig. \ref{Fig6}(f); Path IV: C2 bond with C6 and then C1-C2 bond break to form defect shown by Fig. \ref{Fig6}(g); Path V: C2 bond with C6 and then C5-C6 bond break to form defect shown by defect Fig. \ref{Fig6}(h). Corresponding to these paths, the energy barriers are 5.28 eV, 4.32 eV, 5.77 eV, 5.44 eV and 4.82 eV, respectively. Clearly, the main paths for the defect formation should be path II and path V because they have relatively lower barriers. The time taken by the defect formation via path II and path V is  $2.4\times 10^{54}$ years and $7.8\times 10^{62}$years, respectively. So the lifetime of 6,6,12-graphyne is about $8\times 10^{54}$  years for room temperatures. Even for higher temperature, such as 1000 K, the lifetime is still as long as 7000 years (Tab. \ref{tab1}).

In graphdiyne [Fig.\ref{Fig1}(e)], the possible path for the defect formation path is that the atom nearest to the carbon hexagon C2 and C8 bond together and C1-C2 bond or C7-C8 bond break, as shown in Fig. \ref{Fig6}(i). The barrier is calculated to 3.72 eV. The life time predicted by Eq. \ref{equ3} is $5.4\times 10^{44}$ years at room temperature, but for for higher temperature, such as 1000 K, the lifetime is only 17 years (Tab. \ref{tab1}).

\section{Conclusion}

In summary, the theoretical results show that free-standing single layers of $\alpha$-, $\beta$-, 6,6,12-graphynes and graphdiyne are very stable in temperature range from 300 to 1000 K. So it is possible to prepare free standing graphyne sheets of macrosize, at above room temperature and the device composed of graphyne sheet can work stably even at 1000 K.


%
%

%

\begin{acknowledgments}
The authors are very grateful to acknowledge Professor Qike Zheng for helpful discussions. This work was supported by the National Natural Science Foundation of China under Grant No. 11274073, 11074042 and 51071048, Shanghai Leading Academic Discipline Project (Project No. B107) and Key Discipline Innovative Training Program of Fudan University.
\end{acknowledgments}
%

\end{document}